# Propagation of Rarefaction Pulses in Discrete Materials with Strain-Softening Behavior

E.B. Herbold[1], V.F. Nesterenko[2, 3]

[1]*School of Materials Science and Engineering, Georgia Institute of Technology, Love Manufacturing Building, 771 Ferst Drive, Atlanta, Georgia 30332*

[2]*Department of Mechanical and Aerospace Engineering, University of California at San Diego, La Jolla, California 92093-0411, USA*

[3]*Materials Science and Engineering Program, University of California at San Diego, La Jolla, California 92093-0418, USA*

Discrete materials composed of masses connected by strongly nonlinear links with anomalous behavior (reduction of elastic modulus with strain) have very interesting wave dynamics. Such links may be composed of materials exhibiting repeatable softening behavior under loading and unloading. These discrete materials will not support strongly nonlinear compression pulses due to nonlinear dispersion but may support stationary rarefaction pulses or rarefaction shock-like waves. Here we investigate rarefaction waves in nonlinear periodic systems with a general power-law relationship between force and displacement $F \propto \delta^n$, where $0 < n < 1$. An exact solution of the long-wave approximation is found for the special case of $n = 1/2$, which agrees well with numerical results for the discrete chain. Theoretical and numerical analysis of stationary solutions are discussed for different values of $n$ in the interval $0 < n < 1$. The leading solitary rarefaction wave followed by a dispersive tail was generated by impact in numerical calculations.

Discrete periodic materials with a "normal" power-law relationship between force and displacement, $F \propto \delta^n$ for $n > 1$ (elastic strain hardening), have been shown to support compression solitary waves in periodic granular assemblies [1-11]. In the absence of any initial force applied to grains or when a dynamic force is significantly larger than static precompression these materials are considered strongly nonlinear and recent investigations have considered their use as information carriers and waveguides [3,4,8]. The long wave approximation of the equations of motion for a discrete chain leads to stationary wave solutions traveling in one-dimensional chains or ordered two and three dimensional arrays of particles in the absence of dissipation [1,2]. It was proven in [2] that stationary solitary waves and shock waves should form in discrete materials with a force-displacement relation that stiffens with displacement (for example, power law with $n > 1$). A discrete chain with an interaction law exhibiting general softening behavior (first considered in [12]) supports rarefaction solitary waves and rarefaction shock-like waves [2].

Here we investigate the behavior of stationary rarefaction/release waves in discrete periodic materials composed of point masses and elastically softening links. An exact solution is found for stationary rarefaction solitary waves using the long wave approximation $n = 1/2$ and is compared to numerical simulations of a discrete chain. Stationary rarefaction waves have been investigated in other media like magnetized Hall plasmas and are thought to explain observed anomalous behavior due to a changing electric field [13].

It should be mentioned that in the case of discrete softening materials without tensile strength, the propagation of fracture waves follow directly behind rarefaction pulses if no restoring forces are present.

An elastic or viscoelastic softening behavior (decreasing of elastic modulus with strain) under certain conditions of loading is observed in a wide range of materials from polymer foams [14, 15] and rubber [16] to actin networks in biological tissues [17,18]. In general, the response of materials with "softening" behavior share several common responses under compressive loading: a viscoelastic softening behavior characteristic of configurational changes in polymer chains [15,16,19] or the collapse of cell-wall structures in polymer foams [14, 20-23] followed by a stiffening behavior attributed to the bulk resistance to further deformation. Here, we are concerned only with the simple case where the release wave is from a point along the softening portion of the curve. The interaction law between neighboring particles in a one-dimensional lattice is assumed to be non-dissipative, in experiments it may be realized for foams where softening behavior is due to reversible elastic collapse of cell walls. It can be described by a function of relative particle displacements $\varphi(u_i - u_{i+1})$, where $u_i$ is the displacement of the $i$-th particle from the equilibrium position in the undisturbed chain. The system of discrete equations for a chain of identical particles with a power law potential is

$$\ddot{u}_i = A_{i-1,i}\varphi_{i-1,i} - A_{i,i+1}\varphi_{i,i+1}, \tag{1}$$

where $A$ is an effective stiffness constant and $\varphi_{i-1,i} = (u_{i-1} - u_i)^n$ for $0 < n < 1$. Initial static displacements caused by an external force may also be included in displacement $u_{\mathrm{i}}$. The conditions for propagating stationary rarefaction waves in discrete materials with general 'softening' behavior (and with a specific power-law interaction) are presented in [2].

The long–wave approximation for Eq. (1) is introduced in a way similar to the case of elastic hardening (e.g. particle contact interaction with a Hertzian potential) materials by assuming that the particle diameter, $a$, is significantly less than the propagating wavelength $L$. The result from applying the long wave approximation to Eq. (1) is,

$$u_{tt} = -c_n^2 \left\{ (-u_x)^n + \frac{na^2}{6(n+1)} \left[ (-u_x)^{(n-1)/2} \left( (-u_x)^{(n+1)/2} \right)_{xx} \right] \right\}_x . \quad (2)$$

A derivation of Eq. (2) can be found in [2]. Stationary wave solutions may be found by reducing Eq. (2) to a nonlinear ordinary differential equation corresponding to a stationary waves propagating with speed $V$,

$$y_{\eta\eta} + y - y^{-(n-3)/(n+1)} + y^{-(n-1)/(n+1)} C_2 = 0, \quad (3)$$

for arbitrary values of $n$. In Eq. (3), $y$ is a reduced form of the strain ($\xi = -\partial u / \partial x$),

$$y = \left( \frac{c_n}{V} \right)^{(n+1)/(n-1)} \xi^{(n+1)/2}, \quad (4)$$

where $c_n^2 = Aa^{n+1}$ is a parameter with units of speed and $\eta$ is the normalized coordinate traveling with the speed of the solitary wave $V$,

$$\eta = \frac{x}{a} \sqrt{\frac{6(n+1)}{n}}. \quad (5)$$

Eq. (3) can be rewritten as the equation for a nonlinear oscillator moving in an effective “potential field” $W(y)$,

$$\frac{d^2 y}{d\eta^2} = -\frac{dW}{dy}, \quad (6)$$

where the “potential field” $W(y)$ is defined as following

$$W(y) = \frac{1}{2} y^2 - \frac{n+1}{4} y^{4/(n+1)} + C_3 y^{2/(n+1)}. \quad (7)$$

For an anomalous softening interaction between particles, $0 < n < 1$, rarefaction solitary waves exist when $C_3 = 2C_2/(n+1)$ is bound by [2, 12],

$$\frac{n^2-1}{2}n^{n/(1-n)} < C_3 \le \frac{n-1}{2}\left(\frac{2n}{n+1}\right)^{n/(1-n)}. \tag{8}$$

The value of $C_3$ defines system behavior between weakly and strongly nonlinear regimes, which correspond to the lower and upper bound of $C_3$, respectively.

The sound speed $c_0$ in the chain of particles is found through the linearization of Eq. (2) and is equal to

$$c_0 = c_n\sqrt{n}\xi_0^{(n-1)/2}. \tag{9}$$

A stationary solitary rarefaction wave in a discrete chain of particles may propagate if the chain is initially subject to a constant compression force, $f_0$, which creates an initial strain $\xi_0$.

An analogy can be made that a "particle" in the "potential field" $W(y)$ moves from its initial position ($y_1$, corresponding to $\xi_0$) to the position in the wave corresponding to $y_{min}$(related to $\xi_{min}$) and back to $y_1$. In general, $y_1$ corresponding to the case when a minimum strain is equal zero may be expressed only as a function of the power-law exponent $y_1 = (2n/n+1)^{(1+n)/2(1-n)}$ [2,3]. The total change in the modified strain variable $y$ is relatively large in strongly nonlinear regimes (see Curves 1 and 2 in Fig. 1 (a)) and may be infinitesimally small in the weakly nonlinear regime (see Curve 3) for values of $C_3$ near the lower bound value $C_3$ = -0.1875. In Fig. 1(a), curve 1 corresponds to a case with $n$ = 1/2, $C_3$ = -1/6 (the upper bound of $C_3$ in Eq. (8)) and $y_1 = (2/3)^{3/2}$ is used to show the potential function, Eq. (7), for the special case when the minimum strain is equal to zero.

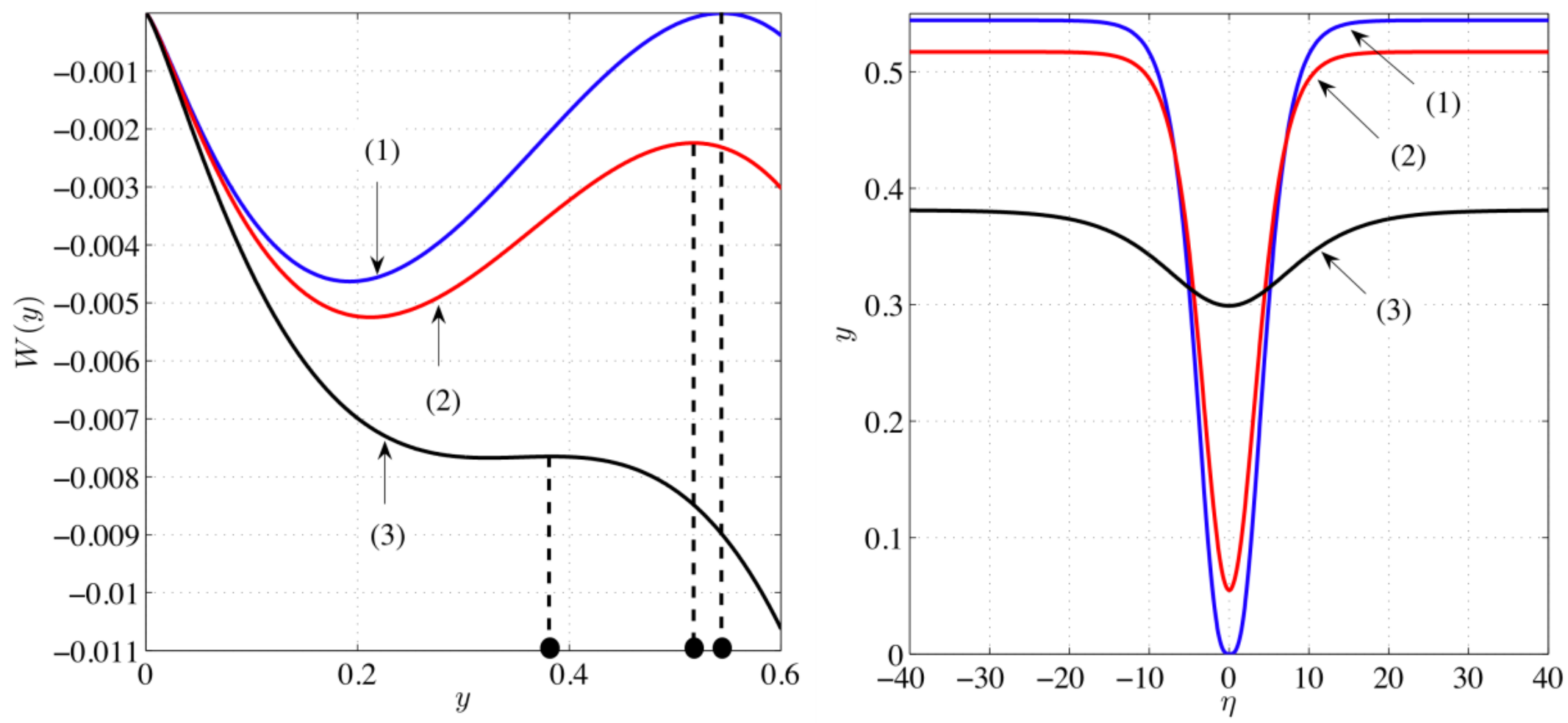

**FIG. 1:** (a) Three curves for different values of $C_3$ in the "potential field" $W(y)$ are plotted for the potential function, Eq. (7), with $n$ = 1/2. Curve (1): $C_3$ = -1/6 and the value of $y$ corresponds to $(2/3)^{3/2}$. Curve (2): $C_3$ = -11/64 and $y_1$ = 0.517. Curve (3): $C_3$ = -0.1869 and $y_1$ = 0.381. (b) Solution of Eq. (3) for different values of $C_2$. The width of the wave is larger for values of $C_2$ that are close to the minimum value. Each curve starts at a different value of $y_1$ (see part (a)). Curve (1): $C_2$ = -2/9, $y_1 = (2/3)^{3/2}$. Curve (2): $C_2$ = -11/48, $y_1$ = 0.517. Curve (3): $C_2$ = -0.2493, $y_1$ = 0.381. All three curves have been shifted horizontally for comparison.

Figure 1 (b) shows the numerical solution of Eqs. (3) and (6) for the three values of $C_2$ (where $C_2 = C_3(n+1)/2$) corresponding to the curves in Fig. 1 (a). The width of the wave increases in the weakly nonlinear case compared to the strongly nonlinear case (compare curves 1 and 2 to curve 3 in Fig. 1 (b)). In the numerical solutions, shown in Fig. 1 (b), the initial locations of the "particles" in the potential field were perturbed by a value of $5e^{-16}$ to induce motion from $y_1$, which is an unstable fixed point. We can see that the shape of the solution for a solitary wave having a minimum strain equal to zero can also be a good approximation for solitary waves with a finite minimum strain (compare curves 1 and 2). It is interesting that the solution having a zero strain at its minimum value corresponds to a singularity in sound speed while also being of similar form to the solutions where such singular behavior of sound speed is absent.

The exact solution for the long wave approximation can be found for the case where the minimum strain is equal to zero and $C_3$ is given by the maximum value in Eq. (8). In the system of reference moving with the wave and centered at the minimum value of strain, the solution can be obtained by integration,

$$\eta = \int_0^y \frac{dy}{\sqrt{-2W(y)}}. \tag{13}$$

An exact solution exists for $n$ = 1/2 and $C_3$ = -1/6 and may be found by substituting Eq. (7) into Eq. (13) giving after integration the following equation,

$$y = \left(\frac{2}{3}\right)^{3/2} \left|\tanh^3\left(\frac{\eta}{3\sqrt{2}}\right)\right|. \tag{14}$$

The corresponding equation for the strain is,

$$\xi = \xi_0 \tanh^4\left(\frac{x}{a}\right). \qquad (15).$$

The exact solution Eq. (15) predicts a symmetric cup-shaped pulse that starts from the initial value of strain, $\xi_0$, decreases to zero and then returns to $\xi_0$. The characteristic length of the pulse based on the properties of the hyperbolic tangent function is equal to $7a$ (for a cut-off of $\xi/\xi_0$=0.98) and does not depend on the amplitude of the solitary wave similar to the case for compressive solitary waves in “sonic vacuum” where $n > 1$ [2]. Equation (14) is shown in Fig. 2, for $n$ = 1/2. In contrast to the strongly nonlinear compression wave, the strongly nonlinear rarefaction wave, Eq. (14) and (15), is not a "compact" solitary wave.

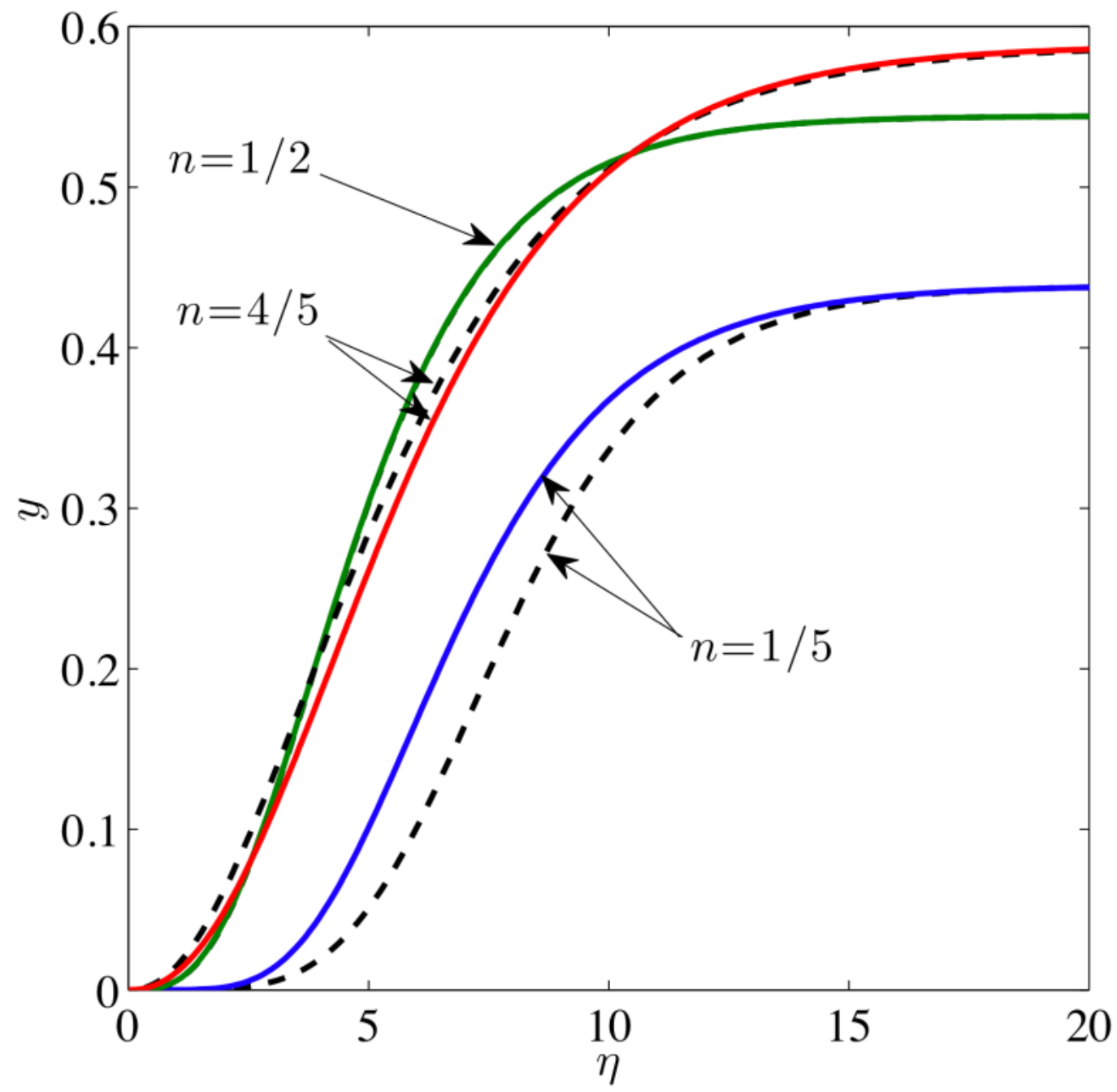

**FIG. 2:** (color online) Equation (14) is compared with the numerical solution of Eq. (3) for three different values of *n*. The amplitudes of the solitary rarefaction waves are exact, but the widths of the pulses appear larger for $n \neq 1/2$.

Fig. 2 shows that the width of the solitary rarefaction wave increases when *n* differs from the value of $n = 1/2$ between 0 and 1. A closed form expression for solitary may be constructed for rarefaction waves for general powers of *n*. Here, the amplitude is equal to the value of $y_1$ and the width of the wave increase for values of *n* in the interval $0 < n < 1$,

$$y = \left(\frac{2n}{n+1}\right)^{(1+n)/2(1-n)} \left| \tanh^{(1+n)/n}\left(\frac{\eta}{\sqrt{3(n+1)/\left(n(1-n)\right)}}\right) \right|. \qquad (16)$$

In Fig. 2 the two remaining comparisons are made between Eq. (16) and the numerical solution of Eq. (3) for $n = 1/5$ and 4/5. The amplitudes of each pulse from Eq. (16) are the same, but the widths of the pulses are slightly underestimated. However, Eq. (16) is a simple closed form approximation of the general form for the long wave approximation.

It is interesting to find the relationships between the phase speed $V$ and the strain $\xi$. The phase speed can be found using the properties of the potential function; $W(y = y_{min}) = W(y = y_1)$ and $\partial W/\partial y|_{y=y_1} = 0$. The speed of the rarefaction wave is [2],

$$V_r = \frac{c_n}{\xi_0 - \xi_{\min}} \left\{ \frac{2\left[ n\xi_0^{n+1} + \xi_{\min}^{n+1} - (n+1)\xi_0^n \xi_{\min} \right]}{n+1} \right\}^{1/2}. \qquad (17)$$

In the case of a solitary rarefaction wave where $\xi_{min} = 0$, Eq. (17) becomes,

$$V_{s,r} = c_n \xi_0^{(n-1)/2} \left( \frac{2n}{n+1} \right)^{1/2}. \qquad (18)$$

The rarefaction wave with a minimum strain ($\xi_{\min}$) equal to zero is a special case where the long wave sound speed represented by Eq. (9) at the point of zero strain is infinite for $n < 1$, but the solitary rarefaction wave speed is finite. We will consider the validity of the rarefaction solitary wave solution in long wave approximation with this singularity point by comparison with results for discrete chain. It should be mentioned that a similar situation exists in the case with strongly nonlinear compressive solitary waves in "normal" material ($n$>1) where there is a periodic solution with a point where sound speed is equal to zero (where initial long wave approximation is not valid) but one hump of each agree very well with solitary wave solution for the discrete system. Numerous numerical investigations using $n > 1$ agree well with the results obtained using the long-wave approximation [2, 5-9]. For example, the classical Hertzian interaction between perfectly elastic spherical particles is a special case of Eq. (4) where $n = 3/2$. Additionally, the ratio of the solitary wave speed, $V_{s,r}$ to the sound speed $c_0$ is equal to $\sqrt{2/(n+1)}$. This ratio is positive $V_{s,r}/c_0 > 1$ for $0 < n < 1$, meaning the solitary rarefaction wave speed is supersonic [2].

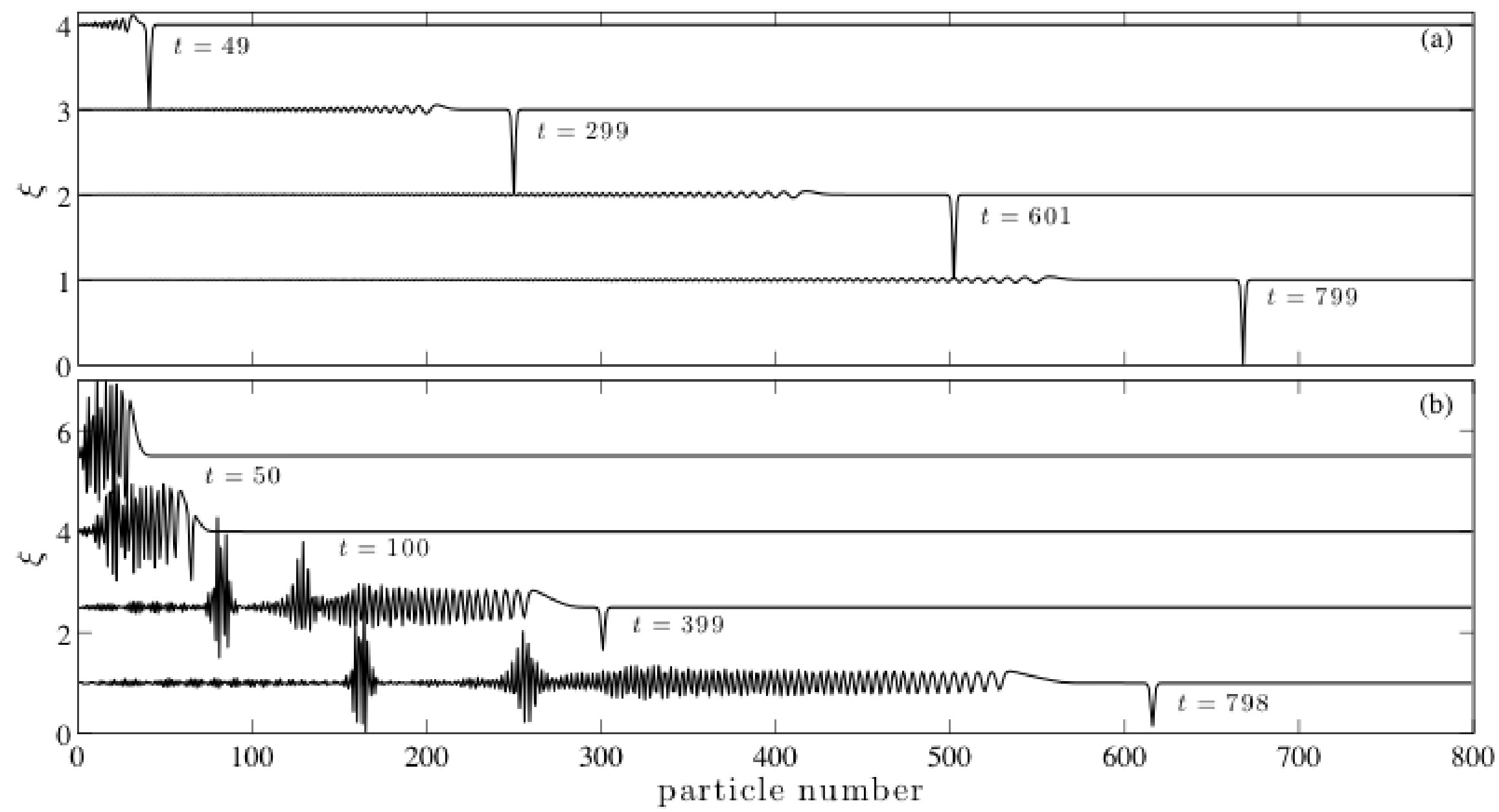

**FIG. 3:** Development of a solitary rarefaction strain wave in a chain of 800 particles compressed with a force of 1 N with a force-displacement relationship with $n$ = 1/2. (a) The initial velocity for the first particle is -1.373 m/s, which has a constant force applied to it. The rarefaction pulse separates from the following oscillatory wave train within approximately 80 particles. (b) The first particle is given and impact velocity of 5 m/s, which also has a constant force applied to it. The compression pulse quickly diminishes due to nonlinear dispersion and the rarefaction pulse immediately preceding it eventually becomes the leading pulse. The strain is offset for visual clarity.

Figure 3 shows the results from simulations using 800 discrete particles where the last particle was fixed and a constant force of $f_0$= 1N is applied to the first particle. The development of a stationary rarefaction pulse is shown by the strain between particles.

The remaining parameters for the simulations were $n$ = 1/2, $a$ = 1m, $A$ = 1 N/kgm$^{1/2}$. At the beginning of the simulation shown in Fig. 3 (a) the first particle is given an initial velocity $v_0$=-1.373 m/s to produce a solitary rarefaction wave with a minimum strain equal to zero, which is a strongly nonlinear rarefaction solitary pulse. The applied force allows the formation of a solitary rarefaction wave in comparison with a release wave where a restoring force is not present. The rarefaction wave moves faster than the oscillatory wave train and separates from it within approximately 80 particles. The minimum of the wave occurs at the 502$^{nd}$ and 250$^{th}$ particles at 601s and 299s, respectively, which gives an average speed of 0.83m/s. Using Eq. (18) with $\xi_0$ = 1 and singularity point at zero strain predicts a speed of 0.82m/s, which agrees within 2%.

In another case an impact velocity of 5 m/s was specified for the first particle in Fig. 3(b), which also has a constant force applied to it. The initial compression pulse quickly diminishes due to nonlinear dispersion and the first rarefaction pulse immediately behind it eventually becomes the leading pulse. It is quite unusual behavior because impact loading is expected to result in compression pulse for "normal" materials. At this impact speed, the resulting minimum strain value of the leading rarefaction pulse is greater and it is equal to $\xi_{min}$= 0.148 at $\xi_0$ = 1. These minimum strain values are shown at particles 616 and 301 at 798s and 399s, respectively, which corresponds to an averaged velocity of 0.79m/s, which is within 2% of the predicted value based on Eq. (17). Eq. (17) resulting from long wave approximation with $c_n$ = 1 and $n$=1/2 gives a speed of 0.78 m/s.

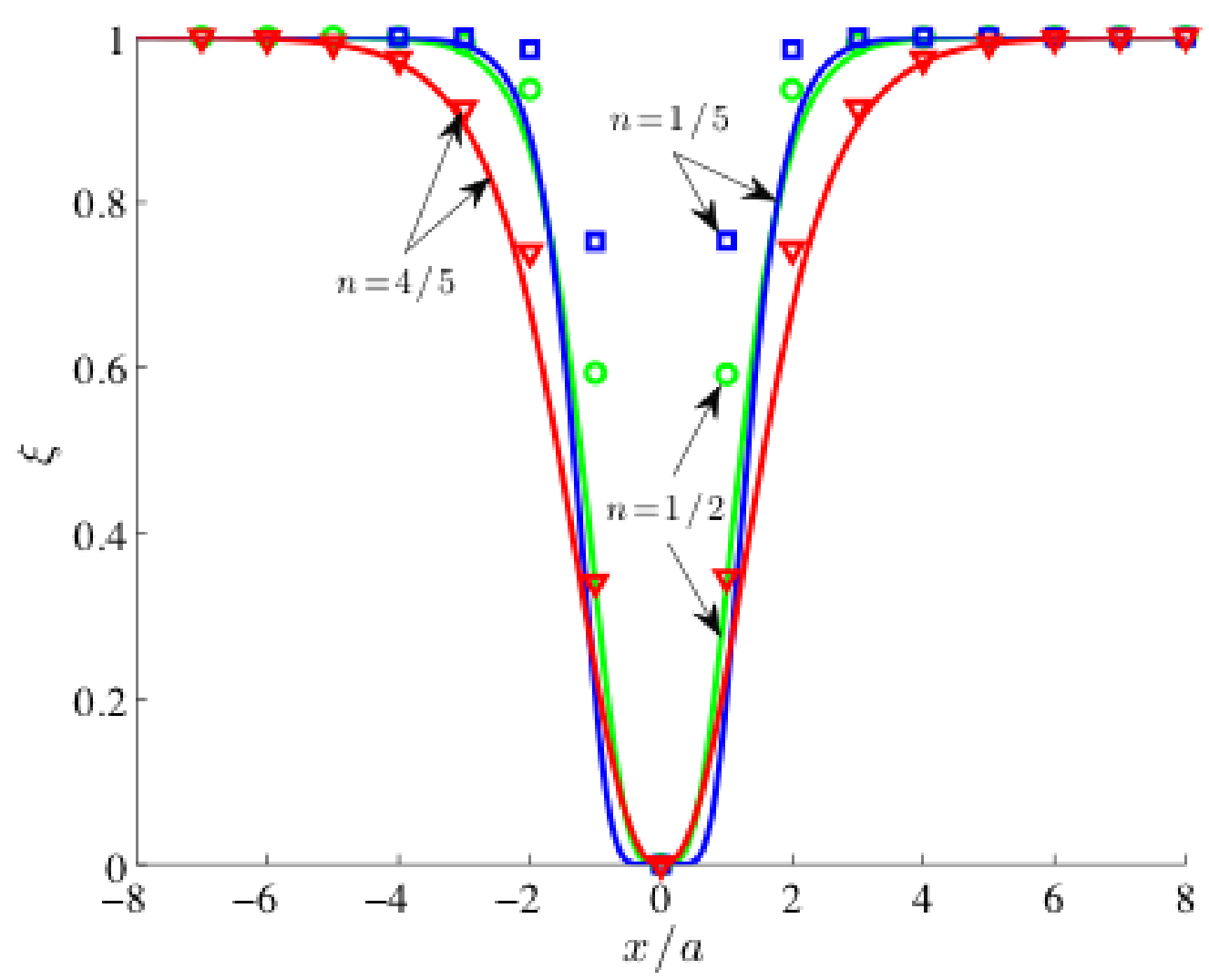

**FIG. 4:** (color online) The strain from discrete simulations with 800 particles is compared to Eq. (16) (scaled from *y* to $\xi$ coordinates) for *n* = 1/5 (blue □), 1/2 (green ○), and 4/5 (red ▿). The remaining parameters for the simulation were the same as for Fig. 3. The width of the solitary rarefaction wave is smaller for the calculated results for each value of *n*.

The agreement between the speed of the leading pulses between the numerical calculation of discrete system of particles and the results from the long wave approximation is excellent for the special case mentioned above where the effective sound speed is infinite for $\xi_{min}$= 0, as well as the case where $\xi_{min}$= 0.148. It is interesting that solitary rarefaction waves may arise by specifying a velocity toward or away from the rest of the chain and suggests different methods to test materials experimentally. This is not possible for strongly nonlinear compression waves in granular media where $n > 1$ due to the absence of a restoring force.

It is important to validate the closed form exact and approximate solutions given in Eq. (14)-(16) by comparison with the discrete simulations for different values of $n$. Figure 4 shows this comparison for $n$ = 1/5, 1/2 and 4/5. The data from Fig. 3 is used for the comparison with $n$ = 1/2 and it is clear that the width of the wave is less than Eq. (15) predicts. However, the discrete calculations show that there are mainly 7 particles comprising the pulse despite this difference. The widths of the pulses are also smaller for the cases where $n$ = 1/5 and 4/5, but agrees the best for the latter. It is interesting that the best fit comes not from the exact solution but from Eq. (16). Fig. (2) showed that Eq. (16) predicted a slightly smaller width of the pulse in comparison to the solution of the long wave approximation, Eq. (3). This suggests that Eq. (16) may more accurately define the shape of the pulses resulting from discrete simulations, though it does not extend to $n$ = 0 or 1.

We investigated rarefaction waves in nonlinear periodic systems with a 'softening' general power-law relationship between force and displacement to understand the dynamic behavior of this class of materials. A closed form expression describing the shape of the strongly nonlinear rarefaction wave is exact for $n$ = 1/2 and describes the shape and width of the pulses resulting from discrete simulations well. The width of the exact solution does not depend on the amplitude of the strongly nonlinear solitary rarefaction wave and it is smallest for $n$ = 1/2 among investigated values of $n$. The agreement between the pulse speed of the waves predicted from the theory and numerical calculations is within 2%. It was shown that the solitary wave speed was supersonic and is inversely proportional to the initial strain of the system. A chain of particles under impact was shown to propagate a rarefaction pulse as the leading pulse in initially compressive impulsive loading in the absence of dissipation.

The authors wish to acknowledge the support of this work by the U.S. NSF (Grant No. DCMS03013220).